\documentclass[letter,twocolumn]{jpsj3}
\usepackage{txfonts}
\usepackage{color}

\title{Impact of Disorder on the Superconducting Phase Diagram in BaFe$_2$(As$_{1-x}$P$_x$)$_2$}

\author{Yuta~Mizukami$^1$\thanks{mizukami@edu.k.u-tokyo.ac.jp}, Marcin~Konczykowski$^2$, Kohei~Matsuura$^1$, Tatsuya~Watashige$^3$, Shigeru~Kasahara$^3$, Yuji~Matsuda$^3$, Takasada~Shibauchi$^1$}
\inst{$^1$Department of Advanced Materials Science, University of Tokyo, Kashiwa, Chiba 277-8561, Japan\\
$^2$Laboratoire des Solides Irradi\'es, \'Ecole Polytechnique, CNRS,
CEA, Universit\'e Paris-Saclay, F-91128 Palaiseau, France\\
$^3$Department of Physics, Kyoto University, Sakyo-ku, Kyoto 606-8502, Japan\\
}

\abst{In many classes of unconventional superconductors, the question of whether the superconductivity is enhanced by the quantum-critical fluctuations on the verge of an ordered phase remains elusive. One of the most direct ways of addressing this issue is to investigate how the superconducting dome traces a shift of the ordered phase. Here, we study how the phase diagram of the iron-based superconductor BaFe$_2$(As$_{1-x}$P$_x$)$_2$ changes with disorder via electron irradiation, which keeps the carrier concentrations intact. With increasing disorder, we find that the magneto-structural transition is suppressed, indicating that the critical concentration is shifted to the lower side. Although the superconducting transition temperature $T_c$ is depressed at high concentrations ($x\gtrsim$0.28), it shows an initial increase at lower $x$. This implies that the superconducting dome tracks the shift of the antiferromagnetic phase, supporting the view of the crucial role played by quantum-critical fluctuations in enhancing superconductivity in this iron-based high-$T_c$ family.}

\begin{document}
\maketitle


In strongly correlated electron systems such as heavy fermions, cuprates, and organic materials, superconductivity often emerges when the antiferromagnetic (AFM) order is suppressed through control parameters such as pressure and chemical composition\cite{Park06, Keimer15}. A striking feature in these materials is that, in several cases, physical properties that deviate from the conventional Fermi-liquid theory (i.e., non-Fermi liquid properties) also appear when the AFM transition is tuned to zero temperature ($T$), suggesting the existence of an AFM quantum critical point (QCP). Although it is widely believed that quantum-critical fluctuations originating from the QCP are closely related to the superconductivity through unconventional pairing mechanisms\cite{Gegenwart08, Broun08}, it remains unclear whether the QCP actually exists inside the superconducting dome. The recently discovered iron pnictides\cite{Hosono15} also exhibit superconductivity in the vicinity of AFM order accompanying tetragonal-to-orthorhombic structural transitions\cite{Dai15}. These magneto-structural transitions can be suppressed by pressure or chemical substitution\cite{Paglione10}, but the quantum criticality is often avoided by a first-order transition in several systems\cite{Dai15}. Among the iron pnictides, Phosphorus(P)-substituted BaFe$_2$As$_2$ is a particularly clean system, and moreover, is unique in the fact that there is growing evidence for the existence of a QCP inside the superconducting dome near the optimal composition\cite{Shibauchi14, Kasahara10, Nakai10, Shishido10, Hashimoto12, Walmsley13}. Although a QCP located at the maximum $T_c$ naturally leads to the consideration that the quantum-critical fluctuations help to enhance superconductivity, there has been no direct evidence against a scenario that it is just a coincidence. A direct test to address this issue is to investigate how the superconducting dome traces when the AFM phase is shifted. However, it has been quite challenging to perform such experiments without changing the carrier numbers or bandwidth, whose effects on the QCP and superconductivity are nontrivial. In fact, in heavy-fermion superconductors, it was highlighted that chemical substitution of dopant atoms may prevent the appearance of quantum criticality altogether\cite{Seo13}.

 Recent advances in the study of the effects of atomic-scale point defects in superconductors using high-energy electron beams allows us to investigate the evolution of the electronic states with increasing impurity scattering in a controlled manner. Through the use of successive electron irradiation, we can perform systematic measurements on a given sample with a gradual introduction of impurity scattering induced by point defects, and without changing the carrier concentration or band width\cite{Prozorov14, Mizukami14}. In general, impurity scattering reduces the $T_c$ in unconventional superconductors, where the suppression rate depends on the gap structures\cite{Balatsky06}. Indeed, the superconducting dome shrinks with the introduction of scattering via chemical substitution in cuprates and heavy-fermion superconductors\cite{Alloul09, Seo15}. Here, we report on the changes of the magneto-structural transition temperature ($T_N$) and superconducting transition temperature ($T_c$) in the $T$-dependence of resistivity $\rho$($T$) with increasing defect density across the entire superconducting dome of BaFe$_2$(As$_{1-x}$P$_x$)$_2$, revealing a monotonic decrease of $T_N$ and highly composition-dependent changes of $T_c$. In particular, the superconductivity initially exhibits an unusual enhancement at low P concentrations with increasing disorder. After irradiation, the superconducting dome exhibits a shift of the optimal composition toward a lower P concentration. This implies that the superconducting dome tracks the AFM phase, supporting the suggested crucial role of quantum-critical fluctuations on the superconductivity in these high-$T_c$ superconducting materials.


\begin{figure*}[t]
        \begin{center}
 	\includegraphics[width=0.9\linewidth]{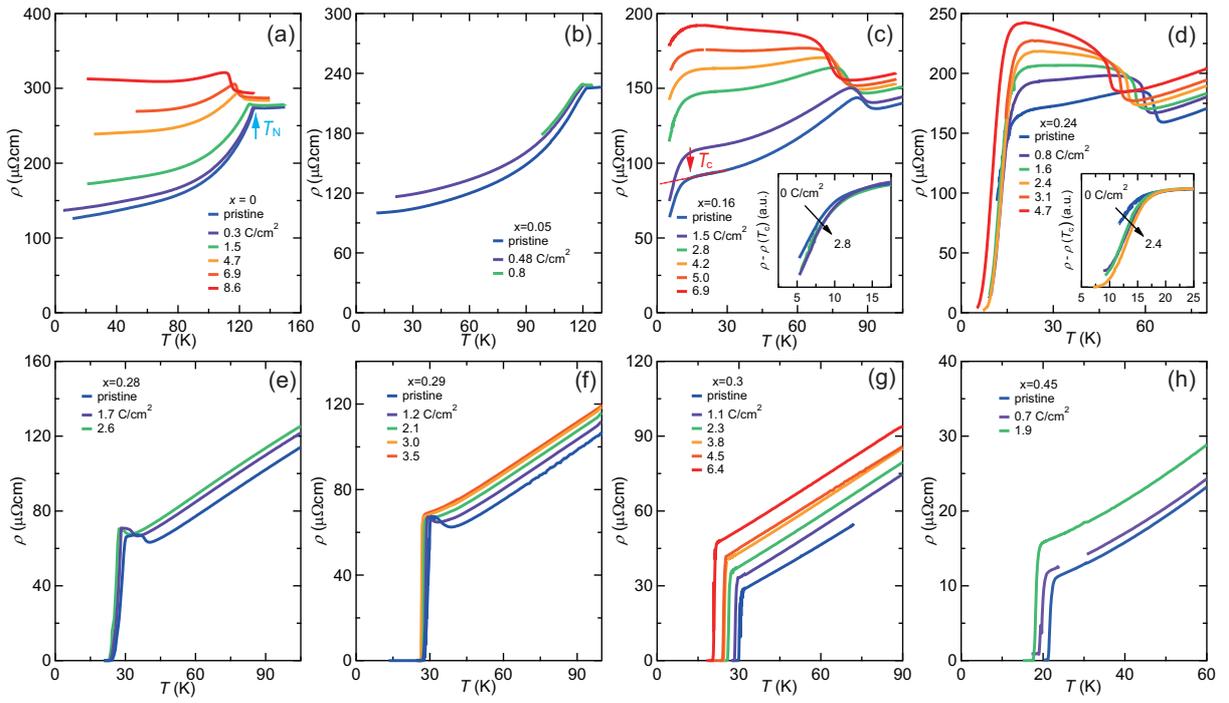}
        \end{center}
 	\caption{
(Color online) (a)--(h) The $\rho$($T$) at different irradiation levels for $x=$0, 0.05, 0.16, 0.24, 0.28, 0.29, 0.3, and 0.45, respectively. The different colors represent different irradiation levels. The blue (red) arrow represents $T_N$($T_c$). The inset in (c) and (d) shows $\rho$($T$) which has been shifted vertically for clarity.}
 	\label{rho}
 \end{figure*}


 Single crystals of BaFe$_2$(As$_{1-x}$P$_x$)$_2$ were synthesized by the self-flux method\cite{Kasahara10}. The quality of the single crystals was confirmed by their sharp superconducting transitions\cite{Kasahara10} and quantum oscillation measurements\cite{Shishido10}. In order to introduce uniform point defects into the BaFe$_2$(As$_{1-x}$P$_x$)$_2$ single crystals, we irradiated the sample with an electron beam with an incident energy of 2.5 MeV, which is far above the threshold energy required for the formation of vacancy-interstitial (Frenkel) pairs\cite{Mizukami14}. The sample was kept at 20 K to prevent defect migration and clustering effects during the irradiation. In order to evaluate the change in $\rho$($T$) with irradiation accurately, we repeated the process of $\rho$($T$) measurement and irradiation on the same crystal without removing the electrodes for each composition. During the irradiation process, $\rho$($T$) was monitored to confirm the increase of $\rho$($T$) induced by defects.

 In Figure\:\ref{rho}(a)--(h), $\rho$($T$) curves are shown at several irradiation levels for $x$ = 0, 0.05, 0.16, 0.24, 0.28, 0.29, 0.3, and 0.45, respectively. For the $x$ = 0 and 0.05 curves, we observe clear kinks at $T_N$ which correspond to 130 K (Figure\:\ref{rho}(a)) and 122 K (Figure\:\ref{rho}(b)) in pristine samples, respectively. Irradiating the sample with electrons causes the observed kinks to split into an upturn and subsequent downturn upon cooling, which is similar to the doping dependence of $\rho$($T$) in BaFe$_2$As$_2$\cite{Ni08, Kasahara10}. The $T_N$ of each composition is monotonically suppressed with increasing irradiation. Above $T_N$, $\rho(T)$ shows an almost parallel shift with irradiation that is caused by $T$-independent impurity scattering, indicating the non-magnetic nature of the point defects. On the other hand, below $T_N$, the change in $\rho(T)$ exhibits some $T$-dependent term whose magnitude becomes larger upon cooling. Such $T$-dependent impurity scattering was also observed with increasing irradiation in the previous $\alpha$-particle irradiation experiment on iron-based superconductor NdFeAs(O,F)\cite{Tarantini10}, and can be understood with the assumption that the magnetic moments of the defects induce Kondo-like scattering, similar to the case of heavy-fermion materials. Although no discernible magnetic moment was observed after irradiation in the paramagnetic state down to 0.1 K in our system\cite{Mizukami14}, this observation indicates that non-magnetic holes created in the AFM networks induce a $T$-dependent scattering process that deserves further investigation to elucidate its origin. For $x$ = 0.16(Figure\:\ref{rho}(c)), and 0.24(Figure\:\ref{rho}(d)), $\rho(T)$ exhibits a reduction at low temperatures due to the onset of superconductivity. Here, $T_c$ is defined as the temperature where $\rho$($T$) starts to drop from the extrapolated linear curve, as indicated in Figure\:\ref{rho}(c). Upon the introduction of disorder, we observe a remarkable feature at several initial stages of irradiation: $T_c$ gradually increases by $\approx$ 1-2\,K with increasing irradiation dosage up to $\approx$ 3\,C/cm$^2$. Although $\rho$($T$) does not reach zero for $x$ = 0.16, the initial increase of $T_c$ can be clearly seen for both $x$ = 0.16 and $x$ = 0.24 when $\rho$($T$) is shifted vertically to compare the $T_c$ at different impurity levels, as shown in the inset of Figure\:\ref{rho}(c) and (d). For $x$ = 0.28(Figure\:\ref{rho}(e)) and 0.29(Figure\:\ref{rho}(f)), which are both near the optimal composition level, $T_N$ decreases with increasing irradiation and even disappears above $\approx$ 1\,C/cm$^2$ for $x$ = 0.29. At the optimal composition $x$ = 0.30 (Figure\:\ref{rho}(g)) and high-P composition $x$ = 0.45 (Figure\:\ref{rho}(h)), no $T_N$ is evident, and we observe the monotonic suppression of $T_c$ with increasing irradiation dosage.

\begin{figure}[t]
        \begin{center}
 	\includegraphics[width=1.0\linewidth]{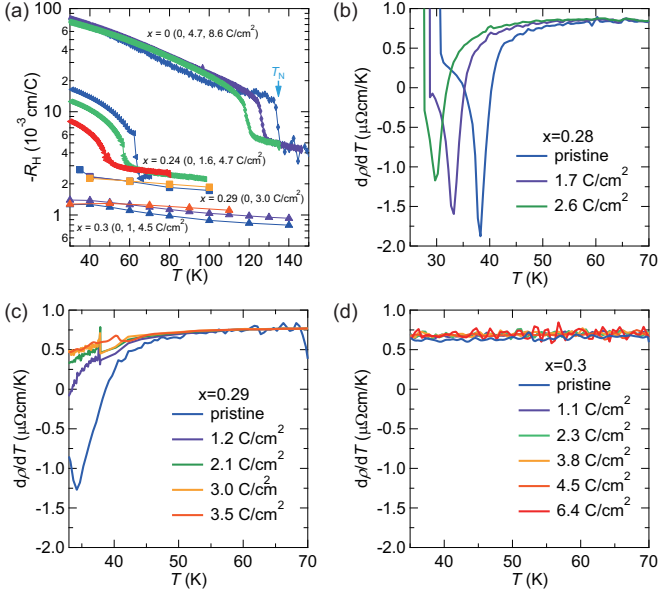}
 	\end{center}
 	\caption{
(Color online) (a) Change of $R_H$($T$) with increasing irradiation for $x=$0 (diamonds), 0.24 (left triangles), 0.29 (squares), and 0.30 (upward triangles). As in Figure\:\ref{rho}, the different colors represent the different irradiation levels. The blue arrow marks the position of $T_N$ where the Hall coefficients experience a jump due to magneto-structural transitions. (b)--(d) Change of d$\rho$/d$T$($T$) with irradiation for $x=$0.28, 0.29, and 0.30, respectively. All curves were obtained by differentiating the $\rho$($T$) data presented in Figures\:\ref{rho}(e)--(g).}
 	\label{hall}
 \end{figure}


 Figure\:\ref{hall}(a) shows the $T$-dependence of the Hall coefficient $R_H$($T$) for $x$ = 0, 0.24, 0.29, and 0.30 at several irradiation levels. In the AFM state, the value of $R_H$($T$) after electron irradiation exhibits a slight change for $x$ = 0.24. The origin of this change may be related to a $T$-dependent scattering process, as in the case for $\rho$($T$). Here, it should be noted that in the paramagnetic state, the change of $R_H$($T$) with irradiation is almost negligible compared to the reported change induced by chemical substitution\cite{Fang09, Kasahara10, Shen11}. This result indicates that electron irradiation does not essentially change the carrier concentration, and mainly introduces impurity scattering. In Figure\:\ref{hall}(b)--(d), the $T$-derivative of the resistivity, d$\rho$/d$T$($T$), is shown for the samples near the optimal compositions $x$ = 0.28, 0.29, and 0.30 at different irradiation levels. The presented data was obtained by differentiating the $\rho$($T$) data shown in Figures\:\ref{rho}(e)--(g). For the $x$ = 0.28 and 0.29 cases, d$\rho$/d$T$($T$) exhibits a sharp dip due to magneto-structural transitions but otherwise remains constant at high temperatures. For $x$ = 0.30, d$\rho$/d$T$($T$) is constant across a wide $T$ range (reflecting the $T$-linear dependence of $\rho$($T$)), and is not affected by the irradiation level. This result demonstrates that electron irradiation does not induce $T$-dependent inelastic scattering, which is in sharp contrast to carrier-doped systems in iron-based superconductors where the change of $T_c$ is concomitant with the drastic evolution of $\rho$($T$). Thus, the change of $T_N$ and $T_c$ with irradiation is not due to a change in carrier number or electron correlations, but is mainly due to an increase in impurity scattering.

\begin{figure}[t]
        \begin{center}
 	\includegraphics[width=1.0\linewidth]{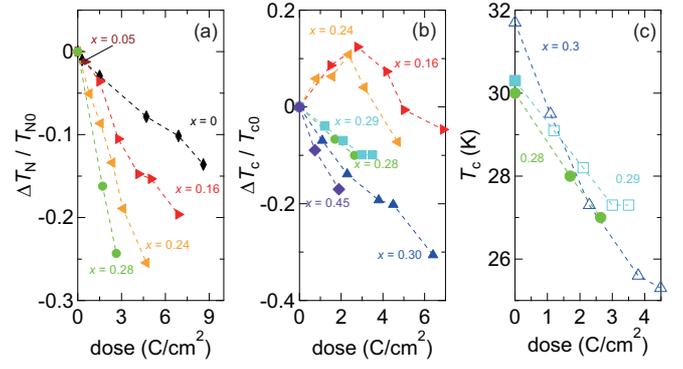}
        \end{center}
 	\caption{
(Color online) (a) Change of $T_N$ from its initial value in a pristine sample $T_{N0}$, and normalized by $T_{N0}$ for $x=$0, 0.05, 0.16, 0.24, and 0.28. (b) Change of $T_c$ from the pristine sample's value $T_{c0}$, and normalized by $T_{c0}$ for $x=$0.16, 0.24, 0.28, 0.29, 0.30, and 0.45. (c) Dose dependence of $T_c$ close to the optimal composition, $x$ = 0.28, 0.29, and 0.3. The green circle, blue square, and dark blue triangle represent the $T_c$ for the $x=$0.28, 0.29, and 0.30 cases, respectively. The open (filled) symbols indicate $T_c$ in the paramagnetic (AFM) state for each composition.}
 	\label{tc}
 \end{figure}

 To see the changes in $T_N$ and $T_c$ caused by irradiation in entire compositions, the dependence of $T_N$ and $T_c$ on the irradiation level is shown in Figure\:\ref{tc}. In Figure\:\ref{tc}(a) (\ref{tc}(b)), the change of $T_N$ ($T_c$) from its pristine value $T_{N0}$ ($T_{c0}$), $\Delta T_N = T_N - T_{N0}$ ($\Delta T_c = T_c - T_{c0}$), is normalized by $T_{N0}$ ($T_{c0}$). Although $T_N$ is reduced by electron irradiation in all compositions, the change of $T_c$ with the irradiation dose displays large composition dependence. For low P concentrations, $x$ = 0.16 and 0.24, as we mentioned earlier, $T_c$ is initially increased and then further levels of irradiation tend to suppress the superconductivity. On the other hand, $T_c$ is monotonically reduced following irradiation for all other compositions, where the magnitude of suppression is larger for higher P concentrations. Figure\:\ref{tc}(c) shows the change of $T_c$ with irradiation dose for compositions near the optimal composition, $x$ = 0.28, 0.29, and 0.30. In pristine samples, the highest $T_c$ is attained for $x$ = 0.30, and irradiation monotonically suppresses $T_c$ for all three compositions. For increased irradiation dose levels, the $T_c$ of $x$ = 0.29 case surpasses the $x$ = 0.30 case above a dose level of 2.0 C/cm$^2$, which can be seen by the crossing of the two curves. Moreover, the $T_c$ for the $x$ = 0.28 case also becomes comparable to the $x$ = 0.30 case around 2.5 C/cm$^2$. These results originate from the fact that the suppression rate of superconductivity becomes gradually larger in cases with high P concentrations, as seen in Figure\:\ref{tc}(b).


\begin{figure}[t]
        \begin{center}
 	\includegraphics[width=1.0\linewidth]{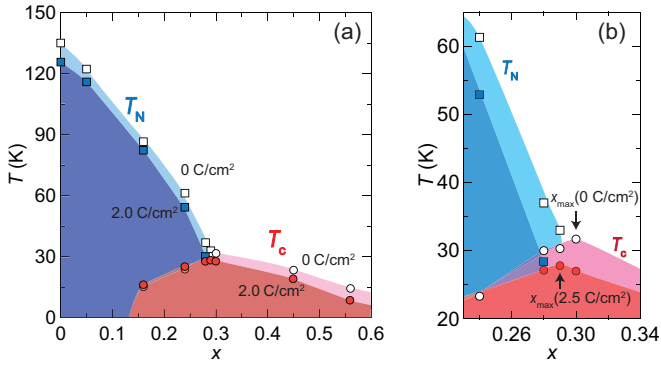}
        \end{center}
 	\caption{
(Color online) (a) The entire phase diagram for 0 and 2.0 C/cm$^2$. The open (filled) squares and circles represent $T_N$ and $T_c$ for 0 (2.0) C/cm$^2$, respectively. The value of $T_c$ for the $x$ = 0.56 sample was determined by magnetization measurements using a Hall-sensor array due to the crystal size. (b) Phase diagram near the optimal composition for for 0 C/cm$^2$ and 2.5 C/cm$^2$. The open (filled) squares and circles represent $T_N$ and $T_c$ for 0 (2.5) C/cm$^2$, respectively. The arrows indicate the compositions where $T_c$ is maximum ($x_{{\rm max}}$) for each irradiation level.}
 	\label{phase}
 \end{figure}

 In Figure\:\ref{phase}(a), we illustrate the phase diagram obtained from the dose dependence of $T_N$ and $T_c$ in Figure\:\ref{tc}. In the phase diagram of the pristine sample, the optimal P concentration coincides with the extrapolated end point of the AFM phase, where the AFM QCP is considered to be located\cite{Shibauchi14}. Here we make the phase diagram for the 2.0 C/cm$^2$ case by interpolating the data points linearly in the dose dependence of $T_N$ and $T_c$. The monotonic decrease of $T_N$ for each composition leads to a shift of the AFM phase. On the other hand, if we look at the change in $T_c$, it displays a strong variation in the magnitude of the suppression, as we mentioned when discussing Figure\:\ref{tc}(b). $T_c$ is increased for low P concentrations, but largely suppressed at high P concentration, as shown by the $T_c$ curve for 2.0 C/cm$^2$ in Figure\:\ref{phase}(a). It is worth noting that there is a clear shift of the optimal composition toward a lower P concentration when we consider the 2.5 C/cm$^2$ phase diagram, as shown in Figure\:\ref{phase}(b).

 Recently, the effect of point defects on $T_c$ on the entire superconducting dome has been reported in hole-doped Ba$_{1-x}$K$_x$Fe$_2$As$_2$\cite{Cho16}. Although the suppression of superconductivity in this system is minimal at optimal doping, and increases away from the optimal doping level, $T_c$ is monotonically reduced at all doping levels for an increasing number of defects. This can be understood in terms of the suppression of superconductivity, which is governed by the magnitude of the gap anisotropy. In the BaFe$_2$(As$_{1-x}$P$_x$)$_2$ case, however, the change of the phase diagram on irradiation is qualitatively different. Here, the superconductivity is enhanced at low P concentrations. Although an increase of $T_c$ due to the introduction of scattering was experimentally reported in Zn-doped LaFeAs(O$_{1-x}$F$_x$)\cite{Li10}, it is not obvious whether Zn substitution introduces only impurity scattering, or whether it involves additional effects such as carrier doping and changes in the lattice parameters. More recently, electron-irradiated FeSe exhibited a slight increase of $T_c \approx$ 0.4 K\cite{Teknowijoyo16}. However, the effect of irradiation in FeSe with very small Fermi energies\cite{Kasahara14} is not well understood, and further investigation is needed to confirm the effect of impurity scattering. Therefore, our result is the first clear observation of a significant increase in $T_c$ merely by impurity scattering.


Indeed, it was already pointed out theoretically that the superconductivity may be enhanced in the AFM regime with the introduction of disorder based on a spin-fluctuation-mediated pairing, if there is competition between the AFM ordering and superconductivity\cite{Fernandes12, Mishra15}. When the enhancement of $T_c$ due to the suppression of AFM order surpasses the reduction of $T_c$ purely from impurity scattering, $T_c$ may be increased as a result of the competing effects. In this scenario, it is expected that the suppression of $T_c$ is largely enhanced when the AFM order is absent. However, we do not observe any significant difference in the suppression rate of $T_c$ between $x$ = 0.28 and 0.29, as shown in Figure\:\ref{tc}(c). Here, magnetism is always present in the $x$ = 0.28 with irradiation case, but $T_N$ disappears rapidly for $x$ = 0.29. This fact indicates that the change of the phase diagram with irradiation cannot be explained merely by the simple competition between AFM order and superconductivity. In fact, the effect of impurity scattering on superconductivity remains, and causes the superconducting dome to shrink. In addition to this effect, if we assume that the entire superconducting dome shifts toward a lower P composition, then we can naturally explain the change of $T_c$ for the entire phase diagram. It should be noted here that the monotonic decrease of $T_N$ naturally leads to the fact that the QCP may also shift its location toward a lower P concentration. Indeed, this is implied by the constant d$\rho$/d$T$, reflecting the fact that the $T$-linear dependence in $\rho$($T$) is extended toward lower temperatures with irradiation for $x$ = 0.28 and 0.29, as shown in Figures\:\ref{hall}(b), and \ref{hall}(c). Here, the constant d$\rho$/d$T$ value is universal between $x$ = 0.28, 0.29, and 0.30. This indicates that the $T$-dependence of $\rho$($T$) in the $x$ = 0.28 and 0.29 cases approaches that of the quantum-critical composition, $x$=0.30, with increasing irradiation. These observations imply that the introduction of disorder results in the shift of the AFM phase toward low P compositions, and that the superconducting dome traces its movement, suggesting that the quantum-critical fluctuations play an essential role in enhancing superconductivity in these iron-based high-$T_c$ superconductors. Such a change of the phase diagram has not been observed in cuprates, which may be related to the fact that the pseudogap temperature does not change significantly with disorder\cite{Alloul09}. 

\begin{acknowledgment}

 We thank H. Kontani, V. Mishra, and R. Prozorov for fruitful discussions. We also thank B. Boizot, O. Cavani, J. Losco, and V. Metayer for technical assistance. This work was supported by the Grants-in-Aid for Scientific Research (KAKENHI) program from the Japan Society for the Promotion of Science (JSPS), and by the ``Topological Quantum Phenomena'' (No.\,25103713) Grant-in Aid for Scientific Research on Innovative Areas from the Ministry of Education, Culture, Sports, Science, and Technology (MEXT) of Japan. The irradiation experiments were supported by the EMIR network, proposal no. 11-10-8071 and no. 15-1580.

\end{acknowledgment}


\begin{thebibliography}{99}

\bibitem{Park06} T.~Park, F.~Ronning, H.~Q.~Yuan, M.~B.~Salamon, R.~Movshovich, J.~L.~Sarrao, and J.~D.~Thompson, Nature {\bf 440}, 65 (2006).

\bibitem{Keimer15} B.~Keimer, S.~A.~Kivelson, M.~R.~Norman, S.~Uchida, and J.~Zaanen, Nature {\bf 518}, 179 (2015).

\bibitem{Gegenwart08} P.~Gegenwart, Q.~Si, and F.~Steglich, Nature Phys. {\bf 4}, 186 (2008).

\bibitem{Broun08} D.~M.~Broun, Nature Phys. {\bf 4}, 170 (2008).

\bibitem{Hosono15} H.~Hosono, and K.~Kuroki, Physica C {\bf 514}, 399 (2015).

\bibitem{Dai15} P.~Dai, Rev. Mod. Phys. {\bf 87}, 855 (2015).

\bibitem{Paglione10} J.~Paglione, and R.~L.~Greene, Nature Phys. {\bf 6}, 645 (2010).

\bibitem{Shibauchi14} T.~Shibauchi, A.~Carrington, and Y.~Matsuda, Annu. Rev. Condens. Matter Phys. {\bf 5}, 113 (2014).

\bibitem{Kasahara10} S.~Kasahara, T.~Shibauchi, K.~Hashimoto, K.~Ikada, S.~Tonegawa, R.~Okazaki, H.~Shishido, H.~Ikeda, H.~Takeya, K.~Hirata, T.~Terashima, and Y.~Matsuda, Phys. Rev. B {\bf 81}, 184519 (2010).

\bibitem{Nakai10} Y.~Nakai, T.~Iye, S.~Kitagawa, K.~Ishida, H.~Ikeda, S.~Kasahara, H.~Shishido, T.~Shibauchi, Y.~Matsuda, and T.~Terashima, Phys. Rev. Lett. {\bf 105}, 107003 (2010).

\bibitem{Shishido10} H.~Shishido, A.~F.~Bangura, A.~I.~Coldea, S.~Tonegawa, K.~Hashimoto, S.~Kasahara, P.~M.~C.~Rourke, H.~Ikeda, T.~Terashima, R.~Settai, Y.~\=Onuki, D.~Vignolles, C.~Proust, B.~Vignolle, A.~McCollam, Y.~Matsuda, T.~Shibauchi, and A.~Carrington, Phys. Rev. Lett. {\bf 104}, 057008 (2010).

\bibitem{Hashimoto12} K.~Hashimoto, K.~Cho, T.~Shibauchi, S.~Kasahara, Y.~Mizukami, R.~Katsumata, Y.~Tsuruhara, T.~Terashima, H.~Ikeda, M.~A.~Tanatar, H.~Kitano, N.~Salovich, R.~W.~Giannetta, P.~Walmsley, A.~Carrington, R.~Prozorov, and Y.~Matsuda, Science {\bf 336}, 1554 (2012).

\bibitem{Walmsley13} P.~Walmsley, C.~Putzke, L.~Malone, I.~Guillam\'on, D.~Vignolles, C.~Proust, S.~Badoux, A.~I.~Coldea, M.~D.~Watson, S.~Kasahara, Y.~Mizukami, T.~Shibauchi, Y.~Matsuda, and A.~Carrington, Phys. Rev. Lett. {\bf 110}, 257002 (2013).

\bibitem{Seo13} S.~Seo, X.~Lu, J-X.~Zhu, R.~R.~Urbano, N.~Curro, E.~D.~Bauer, V.~A.~Sidorov, L.~D.~Pham, T.~Park, Z.~Fisk, and J.~D.~Thompson, Nature Phys. {\bf 10}, 120 (2013).

\bibitem{Prozorov14} R.~Prozorov, M.~Konczykowski, M.~A.~Tanatar, A.~Thaler, S.~L.~Bud'ko, P.~C.~Canfield, V.~Mishra, and P.~J.~Hirschfeld, Phys. Rev. X {\bf 4}, 041032 (2014).

\bibitem{Mizukami14} Y.~Mizukami, M.~Konczykowski, Y.~Kawamoto, S.~Kurata, S.~Kasahara, K.~Hashimoto, V.~Mishra, A.~Kreisel, Y.~Wang, P.~J.~Hirschfeld, Y.~Matsuda, and T.~Shibauchi, Nat. Commun. {\bf 5}, 5657 (2014).

\bibitem{Balatsky06} A.~V.~Balatsky, I.~Vekhter, and J.-X.~Zhu, Rev. Mod. Phys. {\bf 78}, 373 (2006).

\bibitem{Alloul09} H.~Alloul, J.~Bobroff, M.~Gabay and P.~J.~Hirschfeld, Rev. Mod. Phys. {\bf 81}, 45 (2009).

\bibitem{Seo15} S.~Seo, E.~Park, E.~D.~Bauer, F.~Ronning, J.~N.~Kim, J.-H.~Shim, J.~D.~Thompson, and T.~Park, Nat. Commun. {\bf 6}, 6433 (2015).

\bibitem{Ni08} N.~Ni, M.~E.~Tillman, J-Q.~Yan, A.~Kracher, S.~T.~Hannahs, S.~L.~Bud'ko, and P.~C.~Canfield, Phys. Rev. B {\bf 78}, 214515 (2008).

\bibitem{Tarantini10} C.~Tarantini, M.~Putti, A.~Gurevich, Y.~Shen, R.~K.~Singh, J.~M.~Rowell, N.~Newman, D.~C.~Larbalestier, P.~Cheng, Y.~Jia, and H.~H.~Wen, Phys. Rev. Lett. {\bf 104}, 087002 (2010).

\bibitem{Fang09} L.~Fang, H.~Luo, P.~Cheng, Z.~Wang, Y.~Jia, G.~Mu, B.~Shen, I.~I.~Mazin, L.~Shan, C.~Ren, and H-H.~Wen, Phys. Rev. B {\bf 80}, 140508(R) (2009).

\bibitem{Shen11} B.~Shen, H.~Yang, Z-S.~Wang, F.~Han, B.~Zeng, L.~Shan, C.~Ren, and H-H.~Wen, Phys. Rev. B {\bf 84}, 184512 (2011).

\bibitem{Cho16} K.~Cho, M.~Konczykowski, S.~Teknowijoyo, M.~A.~Tanatar, Y.~Liu, T.~A.~Lograsso, W.~E.~Straszheim, V.~Mishra, S.~Maiti, P.~J.~Hirschfeld, and R.~Prozorov, Sci. Adv. {\bf 2}, e1600807 (2016).

\bibitem{Li10} Y.~Li, J.~Tong, Q.~Tao, C.~Feng, G.~Cao, W.~Chen, F.~Zhang, and Z.~Xu, New J. Phys. {\bf 12}, 083008 (2010).

\bibitem{Teknowijoyo16} S.~Teknowijoyo, K.~Cho, M.~A.~Tanatar, J.~Gonzales, A.~E.~B\"{o}hmer, O.~Cavani, V.~Mishra, P.~J.~Hirschfeld, S.~L.~Bud'ko, P.~C.~Canfield, and R.~Prozorov, Phys. Rev. B {\bf 94}, 064521 (2016).

\bibitem{Kasahara14} S.~Kasahara, T.~Watashige, T.~Hanaguri, Y.~Kohsaka, T.~Yamashita, Y.~Shimoyama, Y.~Mizukami, R.~Endo, H.~Ikeda, K.~Aoyama, T.~Terashima, S.~Uji, T.~Wolf, H.~v.~L\"{o}hneysen, T.~Shibauchi, and Y.~Matsuda, Proc. Natl. Acad. Sci. USA {\bf 111}, 16309 (2014).

\bibitem{Fernandes12} R.~M.~Fernandes, M.~G.~Vavilov, and A.~V.~Chubukov, Phys. Rev. B {\bf 85}, 140512 (2012).

\bibitem{Mishra15} V.~Mishra, Phys. Rev. B {\bf 91}, 104501 (2015).

\end{thebibliography}
\end{document}